# High pressure crystal structures of orthovanadates and their properties

Daniel Errandonea*

Departamento de Física Aplicada-ICMUV, MALTA Consolider Team, Universidad de Valencia, Edificio de Investigacion, C/Dr. Moliner 50, Burjassot, 46100 Valencia, Spain

**Abstract:** Pressure-induced phase transitions in orthovanadates have led to interesting physical phenomena. The observed transitions usually involve large volume collapses and drastic changes in the electronic and vibrational properties of the materials. In some cases, the phase transitions implicate coordination changes in vanadium, which has important consequences in the physical properties of vanadates. In this Perspective, we explore the current knowledge of the behavior of $MVO_4$ vanadates under compression. In particular, we summarize studies of the structural, vibrational, and electronic properties and a few illustrative examples of high-pressure research in the compounds of interest are discussed. A systematic understanding of the high-pressure behavior of $MVO_4$ compounds is presented, putting the emphasis on results that could be relevant for practical applications. Recent advances and future challenges in the study of orthovanadates under extreme pressure will be reviewed, along with conclusions that could have consequences for the studies of related oxides. Some ideas on topics that may lead to exciting breakthroughs in the near future will be presented too.

*Corresponding author: daniel.errandonea@uv.es

## I. INTRODUCTION

High-pressure (HP) research has experienced a vast progress in the course of the last two decades [1, 2]. This is a consequence of the combined advancement in instrumentations and experimental techniques as well as in computer simulation methods. Under compression, materials usually undergo a reduction of interatomic distances, which induces important changes in their characteristics, including the vibrational, magnetic, electronic, and mechanical properties. In particular, upon compression, atomic reorganization may be favored inducing therefore phase transitions between different crystal structures [1, 2].

For the family of orthovanadates with formula $MVO_4$, being M a trivalent atom, pressures lower than 10 GPa are enough for inducing structural phase transitions [3]. There are even compounds like $FeVO_4$ [4], undergoing two successive phase transitions below 5 GPa. This feature makes the study of orthovanadates under controlled and quasi-hydrostatic high-pressure conditions quite affordable now-a-days. This has led to the discovery of novel crystal-structure types, which in some cases could be recovered as metastable phases at ambient conditions, showing very different properties than the stable polymorph [3]. Studies under high-pressure conditions have been also useful to test the systematic understanding of the properties of orthovanadates, contributing to deepen the knowledge of their electronic and vibrational properties, which are needed for the efficient expansion of the multiple technologies that have been developed using $MVO_4$ orthovanadates. These materials are currently used as anode materials in Li-ion batteries, as laser-host materials, as thermophosphors, as scintillating detectors for ionizing radiation, in light-emitting diodes, in sensitized solar cells, in supercapacitors, and as photocatalytic materials among many other applications [5 -12]. Part of these applications contributes to the development of environmentally-friendly technologies. In this perspective article, without trying to make an extensive review of the HP behavior of $MVO_4$ orthovanadates, we will comment on the most relevant results of recent studies, their connection will potential applications, and the possible avenues to be explored in the near future.

Among $MVO_4$ orthovanadates, the majority of the studied compounds are those in which the M cation is a rare-earth element [3]. Most of them crystallize in a tetragonal structure which is isomorphic to the crystal structure of zircon (space group $I4_1/amd$) [13]. The only exception is $LaVO_4$, which has a monoclinic structure isomorphic to the crystal structure of monazite (space group P2$_1$/n) [14]. The zircon structure, represented in Fig. 1, is formed by chains of edge-sharing MO$_8$ dodecahedra and VO$_4$ tetrahedra; while in monazite (also shown in Fig. 1) La is coordinated to nine oxygen atoms. In both structures the VO$_4$ tetrahedral units are isolated from equivalent units, being the physical properties of rare-earth vanadates highly influenced by this feature. In particular, it leads to an anisotropic compressibility and to a large electronic band-gap [15]. Zircon- and monazite-type vanadates are isostructural to many other ternary oxides, including silicates, phosphates, chromate, and arsenates [10, 16 – 18]. The larger size of the vanadate ion, in comparison to the silicate and phosphate ions, favors that structural phase transitions occur in the vanadate oxides below 10 GPa, a transition pressure at least 10 GPa smaller than transition pressures in phosphates and silicates. Thanks to it, orthovanadates are more affordable to be studied under compression and can be investigated under nearly hydrostatic conditions, becoming a model for the HP pressure behavior of other ternary oxides.

Multiple HP studies have been carried out in zircon-type vanadates [3]. There is a general understanding, that below 10 GPa, these compounds undergo either a transition to a tetragonal scheelite-type structure (space group I4$_1$/a) [19]; for lanthanide cations with small ionic radii (Sm-Lu); or two a monoclinic monazite-type structure [14]; for lanthanide cations with large ionic radii (Nd, Pr, and Ce). However, it has been recently shown that this systematic can be influenced by non-hydrostatic conditions [20]. In fact, it has been reported that, depending in the degree of non-hydrostaticity, NdVO$_4$ could transform either to the scheelite or the monazite structure [20]. This issue has been little studied in other compounds and deserves to be systematically studied.

Recent progress has been also done recently in the study of mixed lanthanide vanadate compounds like Sm$_x$Nd$_{1-x}$VO$_4$ [21]. Such compounds open the door to the tailoring of the band-gap by changing the relative fraction of Sm and Nd being their photoluminescence a weighted superposition of those of SmVO$_4$ and NdVO$_4$. These mixed vanadate oxides have also studied under high-pressure showing a similar behavior than the end members. One difference with them is the existence of additional metastable phases coexisting with the phases of SmVO$_4$ and NdVO$_4$ [21].

In addition to the detailed characterization of the high-pressure structural behavior of lanthanide vanadates, there have been studies on their optical [13] and vibrational properties [22]. One interesting feature related to the phase transitions reported is the collapse of the band-gap, which in cases reaches up to 1 eV [15, 23]. This phenomenon is connected to the increase of the hybridization between O 2p and V 3d orbitals, which is favored by the rearrangement of the $VO_4$ and $AO_8$ polyhedral units caused by phase transitions. This opens the door to the preparation of materials with smaller band gaps. It also could lead to phenomena like pressure-driven metallization at pressure beyond those reached by high-pressure studies up to date.

The advancement made on the characterization of the HP behavior of orthovanadates includes also the study of the mechanisms driving the phase transitions. Both Raman studies and density-functional calculations have suggested that phonon softening could play a critical role to trigger transitions [20, 24, 52]. This observation has been correlated to that eigenvalues of the stiffness matrix becomes negative, which indicates the presence of a mechanical instability.

In addition to rare-earth vanadates, another group of materials studied under HP are $InVO_4$ [26], $CrVO_4$ [27], $FeVO_4$ [28], and $BiVO_4$ [29, 30]. The main interest in these compounds comes from the fact that they have an optical band gap, which has an optimum value for the non-contaminant production of hydrogen, a zero-emission fuel, from solar energy via the photocatalytic splitting of water [31]. Some on the interesting phenomena observed in these compounds under HP are ferroelastic phase second-order transitions [29, 30] and structural transformations involving a coordination-number increase from four- to six-fold in vanadium. Such structural change triggers a color change of the material and a sudden decrease of the electrical resistivity [26 - 28]; which are related to a band-gap drop of nearly 1.5 eV.

In this perspective article we will focus on describing and analyzing the most relevant achievements on the HP behavior of $MVO_4$ vanadates. However, the aim of the article is not making a comprehensive review, which can be found in the review recently published by Errandonea and Garg [3], but presenting the personal views of the author about future developments, open questions, and possible solutions. Consequently, we will not describe the experimental techniques and computing simulations methods used to study these compounds, because they have been previously broadly described in the literature [32-35]. The rest of the article will be organized as follows. First we will comment on recent achievements on rare-earth vanadates, in particular those related to structural and lattice-dynamics studies. Then we will examine results from mixed vanadate alloys and other $MVO_4$ compounds. After that, we will concentrate on the influence of pressure on the electronic band gap and luminescence properties. After discussing the results available in the literature, we will focus on possible future developments, in special, in those that can contribute to improving technological applications.

## II. RARE-EARTH VANADATES UNDER HIGH PRESSURE

Most rare-earth orthovanadates crystallize in a structure isostructural to that of the mineral zircon ($ZrSiO_4$) [13]. The only exception is $LaVO_4$ which has a crystal structure isomorphic to that of the mineral monazite [14]. $BiVO_4$, $YVO_4$, and $ScVO_4$ also are isostructural to zircon, being $BiVO_4$ dimorphic [36]. All these materials can be prepared in multiple forms, going from large defect-free crystals [37] to nanocrystals [38], which can be optimized for their different applications. As mentioned in the introduction, a schematic representation of the crystal structure of zircon-type $MVO_4$ orthovanadates is shown in Fig. 1a. The crystal structure of zircon belongs to the tetragonal crystal system. being described by space group $I4_1/amd$ [13]. The structure is body centered and contains four formula units per unit cell. It is characterized by axial ratios (c/a) close to $5\sqrt{2}/8$. In the zircon structure, the V and M atoms are at high-symmetry Wyckoff positions. If origin choice 2 used (the most commonly used) to describe the crystal structure [with origin at 2/m and -4m2 at (0, 1/4, 3/8)], the V atoms are at 4a Wyckoff positions (0, 3/4, 1/8) and M atoms are at 4b Wyckoff positions (0, 1/4, 3/8). The oxygen atoms are at 16h Wyckoff positions (0, y, z) with y close to 9/20 (it goes from 0.4279 in $CeVO_4$ to 0.4364 in $LuVO_4$ [3]) and z close 1/5 (it goes from 0.2067 in $CeVO_4$ to 0.1995 in $LuVO_4$ [3]), being the only two atomic coordinates not constrained by symmetry. Considering only the cations, the zircon structure can be seen as sequence of alternate layers of V and M atoms which are shifted one from the other by (0, 1/2, 1/4). As a consequence, the structural framework is formed by chains, of alternating edge-sharing $VO_4$ regular tetrahedral units and $MO_8$ dodecahedra, which are aligned parallel to the c-axis (see Fig. 1a). Each $MO_8$ dodecahedron can be described as made by two interpenetrating $MO_4$ tetragonal disphenoids. The above described chains are linked laterally by edge-sharing dodecahedra (see Fig. 1a). It has been shown that $VO_4$ tetrahedra are very rigid, being little affected by pressure and temperature, dominating the $MO_8$ dodecahedra the thermal expansion and compressibility of zircon-type vanadates.

Under compression, it has been determined that alkaline-earth vanadates undergo a phase transition either to scheelite- or monazite-type structures [3, 20, 21, 39]. These structures are shown in Figs. 1b and 1c. The transition to one structure or to the other is related to the *f*-electron occupation of the lanthanide atom, which correlated with its ionic radii [3]. The systematic high-pressure behavior of this family of compounds in schematically represented in Fig. 2. Compounds from $LuVO_4$ to $NdVO_4$ undergo the zircon-scheelite transition around 5-6 GPa. In contrast, $PrVO_4$ and $CeVO_4$ undergo the zircon-monazite transition taking the structure of $LaVO_4$. However, in compounds closer to the "transition boundary" the structural sequence can be affected by non-hydrostatic conditions, as found in $NdVO_4$, where both the zircon-scheelite and zircon-monazite transition have been observed depending on the pressure medium employed in the experiments [20, 40].

The identification of the crystal-structure of the HP phases has been made by synchrotron powder x-ray diffraction (XRD) experiments. As an example, results on $TbVO_4$ and $CeVO_4$ [41] are shown in Figs. 3 and 4, respectively. The occurrence of the phase transitions at relative low pressure (~6 GPa) favor the measurement of high-quality powder XRD patterns for the HP phase, which lead to an accurate determination of the crystal structures of the HP polymorphs. Both transitions, zircon-scheelite and zircon-monazite, are non-reversible first-order transitions [3]. This fact opens the door to recover the HP polymorphs as metastable phases at ambient conditions. The mechanisms of the transitions have been broadly discussed in the literature and it is beyond the scope of this article [42]. They involve bond switching, being the transitions reconstructive transformations [42]. As can be seen in Fig. 2b, the scheelite structure is also composed by $VO_4$ tetrahedra and $MO_8$ dodecahedra. In monazite, an additional oxygen atom is included in the coordination sphere of the rare-earth atom forming $MO_9$ (see Fig. 2c). As zircon, the scheelite structure is also formed by $VO_4$ and $MO_8$ polyhedra. However, in scheelite the $VO_4$ tetrahedra do not share edges with $MO_8$ dodecahedra, being linked with them by the corners. This gives scheelite the possibility to reduce the "empty" space among polyhedral units by a polyhedral tilting; providing a more dense packing. Such structural arrangement has approximately a 10% smaller volume than zircon (at the same pressure) and a lower enthalpy beyond the transition pressure. In monazite, polyhedral are connected in a similar way than in zircon, However, the polyhedral chains of monazite differ from those of zircon in the fact that they are not as straight making zig-zags. Basically, monazite can be obtained from zircon by a rotation of VO$_4$ tetrahedra combined with a lateral shift of the (100) planes. This gives monazite the flexibility to accommodate large cations (for instance La; being LaVO$_4$ the only MVO$_4$ compound having the monazite structure) or the reduction of volume induced by compression [43] in compounds with cation of similar size than La.

Application of higher pressures leads first to a monoclinic distortion of the scheelite structure [20, 44], triggering a transition to a fergusonite-type structure (space group I2/a) and then to transitions either to post-fergusonite or post-monazite phases [14, 44]. These phases have the common characteristic of having vanadium atoms six-coordinated by oxygen atoms. One of them, the BaWO$_4$-II-type structure is shown in Fig. 1d. We will later discuss the consequences of the vanadium coordination change in the electronic properties of orthovanadates. Fig. 2 shows a schematic representation of the phase diagram of rare-earth vanadates. Notice that YVO$_4$, ScVO$_4$, and zircon-type BiVO$_4$ [45 – 48] behave in the same way than compounds from LuVO$_4$ to NdVO$_4$.

The same behavior could be expected by radioactive PmVO$_4$ [49]. This conclusion is relevant for modeling PmPO$_4$ and other radioactive phosphates as those reported in the framework of the Manhattan project [50]. Notice that the high-pressure behavior is highly sensitive to non-hydrostatic stresses [20, 51, 52]. This is particular relevant for SmVO$_4$, NdVO$_4$, and PrVO$_4$ [20], compounds where the monazite and scheelite structure have similar enthalpies, but not only for them. Indeed, in the case of ScVO$_4$ [52] the occurrence of a non-reversible phase transition from the zircon structure to the fergusonite-type structure has been observed at 6 GPa under highly non-hydrostatic conditions.

Regarding compressibility, all zircon-type vanadates have a bulk modulus ranging from 110 to 150 GPa [3]. A detailed list with linear compresibilities and bulk moduli can be found in Ref. 3. Their compressibility is mainly governed by changes induced by pressure in the MO$_8$ polyhedra. In addition, the linear compressibility has been determined to be anisotropic, with the *c*-axis been approximately 30% less compressible than the *a*-axis. This fact is related to the way that polyhedral are connected [2]. At the phase transition induced by pressure, due to the sudden decrease of the volume, there is a 20% increase of the bulk modulus [2]. Since high-pressure phases are also formed by chains of compressible MO$_8$ (or MO$_9$) units linked by VO$_4$ tetrahedral units, the compression of high-pressure phase is also anisotropic.

The observed phase transitions in MVO$_4$ vanadates can be connected with a few interesting physical phenomena. One is a strong enhancement of the hybridization between V 3d and O 2p orbitals [53], which has consequences in the electronic band gap as we will discuss in one of the next sections. A second phenomenon is the well-known *sp-d* electron and *f*-electron delocalization induced by pressure in lanthanides [54, 55]. Another fact is the triggering of a violation of the elastic stability conditions in the zircon phase at a pressure just above the transition pressure [44]. In particular, according with density-functional theory calculations, above the transition the Born stability criteria are violated in the zircon structure [56]. Thus computing simulations support the experimental finding not only with thermodynamic arguments (the free energy of zircon becomes higher than that of the HP phases of MVO$_4$ compounds), but also with mechanical arguments. In particular, computing simulations have played a key role in the microscopic understanding of phase transitions [3]. In order to test theoretical predictions, it would be useful to perform in the future Brillouin scattering measurements under high-pressure, which up to now have been only made for ErVO$_4$ at ambient conditions [57], but can be implemented for HP experiments using diamond-anvil cells [58]. A forth event related with phase transition in zircon-type vanadates, but not the last, is the presence of vibrational modes with negative pressure coefficients. In particular, the existence of a soft silent mode, related to rotations of the VO$_4$ tetrahedra, with a frequency that becomes zero around the transition pressure [44, 59]. All these facts show that the study of zircon-type vanadates is significant not only because of their applications, but also for fundamental research. Recently, neutron inelastic-scattering experiments validated calculated phonon dispersion for YVO$_4$ [60]. The same study showed that the calculated pressure dependence of phonon modes in the zircon phase shows unstable modes and violation of the Born stability criteria at high pressure, which agree with previous studies suggestion that phase transitions in zircon-type vanadates may be related to phonon instabilities at high pressures [20, 24, 53]. A similar study has been performed only for LuVO$_4$ [61]. The extension to other vanadates of neutron-inelastic scattering experiments combined with lattice-dynamics calculations would contribute to understand better the transition mechanism of high-pressure phase transitions in vanadates.

Raman spectroscopy is the ability to provide a great amount of information. It can be used to characterize the elastic and vibrational properties giving fingerprints of the occurrence of phase transitions. Raman studies on vanadates, in many cases, triggered the rest of high-pressure studies described in the work. According to group theory, the zircon structure has twelve Raman-active modes ($2A_g + 4B_{1g} + 1 B_{2g} + 5E_g$). They have accurately identified from polarized Raman experiments carried out in single crystals [62]. The modes can be assigned either to internal vibrations of the $VO_4$ tetrahedron and vibration that involve the movement of lanthanide atoms and $VO_4$ tetrahedra as rigid units [60]. Usually ten or eleven Raman modes are measured [62, 63, 64] being one $E_g$ mode difficult to be detected because of it has a weak Raman scattering cross-section and overlaps with one of the $A_{1g}$ modes. The twelve modes have been only measured in $NdVO_4$ [23]. In order to illustrate changes induced by pressure in the Raman spectrum we show results measured in $TbVO_4$ [59]. In this diamond-anvil cell experiments, nine Raman modes are detected. The three strong modes at high-frequencies are internal modes of the $VO_4$ tetrahedron. The typical phonon gap of the Raman spectrum of zircon-type vanadates can be also seen in the figure. In the figure, it can be also seen the important changes that occur in the Raman spectrum at the successive phase transitions. For instance, at the zircon-scheelite transition there is abrupt decrease in the frequencies of the internal modes of the $VO_4$ tetrahedron, which is consistent with a weakening of the force constants associated to these modes. In Fig. 6 we show the pressure dependence of the Raman modes in $TbVO_4$, which is representative of the behavior of the rest of the compounds of the zircon-type vanadate family. The most interesting feature of bond phases is the presence of one mode in each phase in which the frequency decrease with pressure. These are the modes shown in red in the figure. The presence of these modes is an indication of the presence of deformations in the crystal structure which favor the occurrence of structural phase transitions [65, 66].

An interesting area to explore in the future is the behavior of rare-earth vanadates under low-temperature and high-pressure conditions. It has been known for a decade that $TbVO_4$ undergoes a structural phase transition below 33 K [67]. This transition is induced by a cooperative Jahn-Teller effect and it is related to a magnetic transition to an antiferromagnetic state [68]. It has been observed not only in $TbVO_4$ but also in other members of the family [69, 70]. By single-crystal XRD it has been determined that the transition is from the tetragonal zircon structure to an orthorhombic structure which lowers the space group symmetry from $I4_1/amd$ to $Fddd$. The transition occurs without volume changes associated to it. In addition, the space group of the low-temperature phase (Fddd) is related to the space group of the high-temperature phase (I4$_1$/amd) by a group-subgroup relationship. Both facts and the existence of soft-phonon modes point towards a second-order nature [71, 72] for the low-temperature transition of rare-earth vanadates [70]. It would be very interesting to explore in the future how the low-temperature phase transition is affected by compression; in particular, considering that under compression, the Jahn-Teller distortion, driving the low-temperature transition, is expected to persist under HP, inducing structural changes, in order to minimize the energy of the system [73]. Such studies can be performed with current techniques by combining low-temperature and high-pressure neutron and single-crystal diffraction [74, 75]. Another interesting phenomenon to study by the combination of low-temperature and high-pressure are the anharmonic coupling of Raman-active phonons with the electronic states [76]. These modes have been proposed to have a soft-mode behavior causing the low-temperature transition to occur.

Results obtained from zircon-type vanadates can be used to model the high-pressure behavior of silicates [77], phosphates [17], germanates [78], selenates [79], arsenates [80], chromates [80], and any other oxide isostructural to zircon. ZrSiO$_4$ undergoes the zircon-scheelite transition (also known as zircon-reidite transition) at 10 GPa at elevated temperature [81], but at room temperature the phase transition occurs at ~ 22.4 GPa, after zircon transforming into a low-symmetry high-pressure phase acting as a structural bridge between the zircon-type and scheelite-type structures [81]. On the other hand, zircon-type phosphates undergo phase transitions at pressure higher than 12 GPa [82]. These pressures double at least the transition pressure of vanadates. This fact can be explained using crystal chemistry arguments [10]. The ionic radius of tetra-coordinated V$^{5+}$ is 0.35 Å, the ionic radius of tetra-coordinated P$^{5+}$ is 0.17 Å, and the ionic radius of tetra-coordinated Si$^{4+}$ is 0.26 Å. The increase of the cation size will favor the overlap of its orbitals with those of other atoms. Thus, as a first approximation, a vanadate can be seen as a phosphate or a silicate suffering a "chemical pressure" which has a similar effect on the crystal structure that external high pressure. This does not only explain the occurrence of phase transitions at lower pressure in vanadates than in phosphates and other oxides, but also allows to use vanadates as model for the other compounds. This is because vanadate can be studied more easily under controlled hydrostatic conditions. In particular, the existence of transitions to structures with six-coordinated vanadium at pressure of the order of 30 GPa, suggest that six-coordinated silicates and phosphates can be formed in the lower mantle at a depth of around 1000 Km (~40 GPa). In special, the amorphization of $ZrSiO_4$ around 30 GPa [81] could be related to the frustration of this kind of transitions. The study of this subject could be triggered by the results reported for orthovanadates.

### III.   MIXED RARE-EARTH VANADATES UNDER HIGH PRESSURE

Mixed rare-earth vanadates (for instance $Lu_xY_{1-x}VO_4$) have recently attracted attention because they give the possibility to fine tune the electronic and thermal properties optimizing the materials for different applications [83 - 86]. An example of it can be seen in Figure 7, where we reproduce the luminescence spectrum at ambient conditions of $Sm_xNd_{1-x}VO_4$ solid solutions with different compositions [21]. The behavior of the compounds evidently falls into two different behaviors. For large concentrations of Nd the luminescence bands due to $Nd^{3+}$ are quenched because of the presence of $Sm^{3+}$, being photoluminescence detectable only for pure $NdVO_4$. In contrast, for large concentrations of Sm, photoluminescence is only related to $Sm^{3+}$, with the intensity growing as the Sm concentration is enhanced.

Compounds with mixed compositions can have a different behavior that their end members when studied under high-pressure [87, 88]. For instance, it has been observed that $Tb_{0.5}Gd_{0.5}PO_4$ does not undergo a transition from zircon to scheelite or monazite as expected for $TbPO_4$ and $GdPO_4$ [17]. In contrast, this solid solution undergoes a transformation to an anhydrite-type phase which is a crystal structure not expected to be observed when compressing zircon according to crystal chemistry arguments [10]. On the other hand, it is well-known that partially substituting a cation by a different atom affects both the transition pressures and compressibility of zircon-type oxides [89]. A similar, "atypical" behavior could be expected for solid solutions of rare-earth vanadates. However, very little has been done on these compounds under pressure.

To the best of our knowledge, only very recently $Sm_{0.5}Nd_{0.5}VO_4$ [21] and $Eu_{0.1}Bi_{0.9}VO_4$ [90] have been studied under compression. Results from powder XRD experiments on $Sm_{0.5}Nd_{0.5}VO_4$ [21] are summarized in Fig. 8. The behavior of this solution is not exactly the same than the behavior of $SmVO_4$ and $NdVO_4$ [40, 91]. In particular, $Sm_{0.5}Nd_{0.5}VO_4$ first has a phase transition to a mixture of two polymorphs, one isomorphic to scheelite and the other isomorphic to monazite. This could be related to the separation of the two compounds (of SmVO$_4$ and NdVO$_4$); however, this hypothesis is ruled out by the subsequent transition to a single scheelite-structured phase and third transition to a fergusonite-type structure around 16 GPa, a pressure quite lower than the scheelite-fergusonite transition pressure in SmVO$_4$ and NdVO$_4$ [20, 90]. In addition, a mixture of zircon and scheelite is found after pressure release. The coexistence of the monazite- and scheelite-type phase has been explained by enthalpy calculations [21] showing that for the solid solution both phases has a similar enthalpy in the 8 - 14 GPa pressure range.

In the case of Eu$_{0.1}$Bi$_{0.9}$VO$_4$ [91] also a very interesting behavior has been observed. This compound first undergoes a second-order ferroelastic-paraelastic fergusonite-sccheelite transition at 1.9 GPa mimicking the behavior of BiVO$_4$ [30]. However, at 16 GPa a second transition has been reported to a structure not observed before in any MVO$_4$ vanadate [91]. The second phase transition is first-order in nature, being a monoclinic structure (space group P2$_1$/n) proposed for the second high-pressure phase [91]. This phase is expected to have a smaller band gap than BiVO$_4$, being probably smaller than 2 eV. The results on Sm$_{0.5}$Nd$_{0.5}$VO$_4$ [21] and Eu$_{0.1}$Bi$_{0.9}$VO$_4$ [91] show that solid mixtures could open the door to a different high-pressure than in single composition MVO$_4$ vanadates, leading to unexpected phenomena. This suggests that the study under HP of compounds with mixed compositions is one of the interesting avenues to be explored in the future. Of interest could be not only compounds with mixed ternary cation composition, but also with mixed pentavalent atoms, for instance TbV$_{1-x}$P$_x$O$_4$ and similar compounds [92]. It would be quite fascinating also to study Gd$_{1-x}$La$_x$VO$_4$ [93]. Depending on La concentration the structure of this compound is expected to change from zircon to monazite. This may lead to interesting results for intermediate compositions under high pressure.

## IV.   OTHER ORTHOVANADATES UNDER HIGH PRESSURE

Great interest has been recently received by other MVO$_4$ vanadates, including InVO$_4$, FeVO$_4$, and CrVO$_4$. The interest on them comes from their potential technological applications. They include the use as anode materials in Li-ion batteries [94], superionic conductors [95], photocatalytic materials [96], magnetic materials [97], and in electrochemical applications [98]. FeVO$_4$ has a rich polymorphism with several crystal structures reported in the literature [99]. The ambient-conditions structure is triclinic (space group $P\bar{1}$) [99]. It is represented in Fig. 9. This structure is also shared by AlVO$_4$. It consists of a framework of FeO$_6$ octahedra and FeO$_5$ trigonal bipyramids chains, which are linked by VO$_4$ tetrahedra. In the case of InVO$_4$ and CrVO$_4$, the two compounds are isomorphic, sharing the structure also with TlVO$_4$ [100]. The crystal structure of these compounds is shown in Fig. 9. It is orthorhombic, being known as CrVO$_4$-type, and contains VO$_4$ tetrahedra and MO$_6$ octahedra as building blocks. The structure can be described as chains of MO$_6$ octahedra running along the *c*-axis, which are connected via VO$_4$ tetrahedra. In both the triclinic and orthorhombic structures of this group of compounds, VO$_4$ tetrahedra are isolated.

Great progress on the understanding of the properties the compounds discussed in this chapter has been made on studies of thin films [101] and nanoparticles [102]. High-pressure studies were first performed in the early sixties. In particular, it was reported that new polymorphs could be synthesized from FeVO$_4$ and CrVO$_4$ at 750ºC and 6.5 GPa [103]. However, the crystal structure of these polymorphs was not solved. These compounds have been extensively studies under variable temperature [104]. Recently, InVO$_4$ [11, 26, 105, 106], FeVO$_4$ [4, 28], and CrVO$_4$ [27] have been studies under high-pressure conditions.

In InVO$_4$, a transition from the orthorhombic stable structure to a monoclinic structure has been found to occur near 8 GPa. The transition was detected both by Raman spectroscopy and XRD [26] and it is also supported by density-functional theory calculations [105]. The changes caused by pressure in powder XRD patterns can be seen in Fig.10. The HP phase is monoclinic and involves a coordination change in vanadium. Its coordination number changes from four to six. This can be seen in Fig. 10 where the low- and high-pressure phases are shown for comparison. The HP phase is monoclinic and can be described by space group P2/c [26]. The structure is isomorphic to that of InTaO$_4$ [107] and can be considered as a wolframite-type structures [108]. The existence of structures with V atoms in six-coordination occurs at much lower pressure than in lanthanide vanadates. Other finger print of InVO$_4$ and related oxides is their larger compressibility, with bulk moduli below 70 GPa [4, 11, 26, 27, 28, 105, 106]. However, after the phase transition, due to a huge volume collapse; a relative volume change of 20%, the density of the compounds is considerably increased and the bulk modulus becomes comparable to that of zircon-type vanadates. Analogous results to those above described for InVO$_4$ has been also reported for CrVO$_4$ [27]. The changes induced by pressure in the crystal structure drive important changes in the electronic properties which will be discussed in the section devoted to electronic properties.

In the case of triclinic $FeVO_4$, the high-pressure behavior than in $InVO_4$ and $CrVO_4$, but still some similarities can be found with these compounds. A combination of high-pressure single-crystal x-ray diffraction measurements and density functional-theory calculations [4] have shown that $FeVO_4$ has an extremely reach polymorphism under compression. In particular, in this vanadate the first transition takes place at 2.1 GPa being the structure of the HP polymorph also triclinic, but having both Fe and V cations in an octahedral coordination. In addition, below 10 GPa two additional phase transitions have been reported [4, 10]. Interestingly, the third HP phase is isomorphic to wolframite as the HP polymorphs of $InVO_4$ [26] and $CrVO_4$ [27]. Thus, systemically all this group of compounds systematically transforms into a wolframite-type structure in which V is in six-coordination, in which are expected to have a much smaller electronic band-gap, which is supported by color changes observed in the samples at the phase transitions [4, 26, 27] and by optical and resistivity measurements that we will discuss in the next section. Based upon crystal chemistry arguments and the high-pressure systematic behavior of related oxides [10] it would not be surprising that also $AlVO_4$ and $TlVO_4$ transform under compression (directly or by means of successive transitions) into a wolframite-type polymorph. Therefore it would be interesting to study these compounds under HP in the future. Another interesting subject to be explored is the existence of post-wolframite phase. Such phase has been discovered in $InNbO_4$ and $InTaO_4$ [107, 109]. It is known to have a considerable smaller band gap than the wolframite-type phase. Since in the case of $InVO_4$ and $CrVO_4$ [28, 101] the wolframite phases has band gaps in the near infrared, it would be interesting to investigate if pressure could induce metallization in them by mean of an electron transfer from O 2p states to transition metals 3D states [110].

A final subject of interest for future studies is the behavior of vanadates of mixed concentrations related to the compounds discussed in this section. In particular, $Fe_{1-x}Bi_xO_4$ [111], $FeVO_4$-$CrVO_4$ [112, 113], and $CrV_{1-x}P_xO_4$ [114] solid solutions have been successfully synthesized. In $Fe_{1-x}Cr_xVO_4$ for Cr concentrations corresponding to x < 0.175), the crystal structure is the triclinic structure of $FeVO_4$, for Cr concentrations corresponding to x > 0.25, the crystal structure is the orthorhombic structure of $CrVO_4$. However, for 0.175 < x < 0.25, a monoclinic phase (C2/m space group) emerges. The study of these solutions could be very interesting since it could lead to unexpected results, helping to understand the HP behavior of this family of vanadates.

## V. ELECTRONIC PROPERTIES

The accurate knowledge of the electronic properties of $MVO_4$ vanadates is crucial for the developing of their technical applications. Several studies have been carried out to characterize these properties in rare-earth vanadates; they have been carried out at ambient conditions and under high-pressure [15, 23, 44, 115, 116, 117]. The common understanding is that the top pf the valence band and the bottom of the conductions band are dominated by oxygen 2p states and vanadium 3d states. The fact that band structure near the Fermi level originates mainly from the orbitals of the vanadate ion, make most rare-earth vanadates to have band gaps close to 3.8 eV. However, for compounds like $CeVO_4$ and $LaVO_4$ [115, 116] there is a contribution from 4d and 5f states of the lanthanide to the bottom of the conduction band, which leads to a reduction of the band-gap energy. These basic ideas describing the electronic structure of lanthanide vanadates are schematically represented in Fig. 11. They also apply to compounds like $YVO_4$, whose outer electrons (4d and 5s) do not contribute to the states near the Fermi level, having a band-gap close to 3.8 eV [15, 115,118]. In the case of $BiVO_4$, the coupling between Bi 6s and O 2p moves upward the top of the valence band and the coupling between V 3d, O 2p, and Bi 6p, lowers the bottom of the conduction band, reducing the band-gap energy to values close to 2.5 eV [115, 119]. In the cases of $InVO_4$, $CrVO_4$, and $FeVO_4$, the contribution of In, Cr, and Fe states to conduction and valence band lowers the band-gap energy to values close to 3.2, 2.7, 2.1 eV, respectively [27, 105, 106, 120, 121, 122].

High-pressure studies have helped in the understanding of the electronic properties of orthovanadates. In particular, they confirm that in most rare-earth vanadates the electronic bands near the Fermi level are dominated by O 2p and V 3d states. When pressure is increased, V-O bond distances decrease. As a consequence, crystal filed is enhanced, increasing the repulsion between bonding and antibonding states [15, 23, 46]. As a consequence, the bottom of the conduction band moves faster toward higher energies than the top of the valence band, causing a gradual band-gap opening. This has been confirmed by optical-absorption measurements, like those shown in Fig. 12 for $YVO_4$. In the figure, it can be seen a gradual blue-shift of the fundamental absorption edge under compression. A similar behavior has been found in many zircon-type vanadates [15, 23, 46], having the band-gap energy a linear pressure coefficient, as shown in Fig. 12 for YVO$_4$. The coefficients run from 5 to 20 meV/GPa; depending on the compound.

High-pressure studies have also shown that the contribution of states from the trivalent atom to the conduction band lead not only to a reductions of the band-gap, as described above, but also to a very different pressure behavior of the band-gap energy. This can be clearly seen by comparing in Fig. 12 the behavior of LaVO$_4$ with that of YVO$_4$. The different behavior of the band-gap energy in LaVO$_4$ comes from the contribution of La states to the bottom of the conduction band. It causes that while in other rare-earth vanadates under compression the bottom of the conduction band goes up in energy, the contribution of La states makes it to go down [116]. This is translated in a red-shift of the band-gap of LaVO$_4$ as can be seen in Fig. 12. By analogy a similar behavior can be expected for CeVO$_4$.

At the phase transition induced by pressure, in all the studied compounds, an abrupt decrease of the band-gap energy has been found [15, 23, 44, 116]. This is shown for YVO$_4$ and LaVO$_4$ in Fig.12. At the transition, due to the volume collapse and the decrease of symmetry there is an enhancement of orbital hybridization, which cause the observed collapse of the band gap. In addition to that, also the luminescent properties of rare-earth vanadates are strongly affected by pressure [116, 120]. Among them it can be remarked the pressure-enhanced light emission of Er-doped GdVO$_4$, which is connected to changes induced by pressure in the crystal structure [123]. Luminescence properties are also affected by compression in LaVO$_4$ [116]. In particular, drastic changes take place when the phase transition from monazite to the BaWO$_4$-II structure occurs. These changes are related to changes in the coordination number of vanadium ions and in the local sites of La$^{3+}$.

In the past, it has been suggested that zircon-type vanadates were good candidates for the occurrence of pressure-induced metallization [124]. However, such phenomenon has not been observed yet in experiments. In the case of CeVO$_4$, the electrical resistance undergoes a large drop in resistance after a phase transition around 10 GPa, indicating a substantial decrease of the band gap [125]. However, the value of the resistivity after such drop indicates that CeVO$_4$ behaves as a narrow gap semiconductor and not as a metal. The low resistivity of the HP phase of CeVO$_4$, is probably related to the formation of sharp donor levels, which originate from the creation of vacancies due to the migration of oxygen atoms to defects created after the phase transition [126].

In spite that InVO$_4$ and CrVO$_4$ [27, 106] have smaller band gaps that rare-earth vanadates, the pressure dependency of the band-gap energy is similar than in lanthanide vanadates. Results for CrVO$_4$ are shown in Fig. 13. Both experiments and DFT calculations give a qualitative similar behavior. CrVO$_4$ is a direct band-gap material, with the top of the valence band and the bottom of the conduction band at the Γ point of the Brillouin zone. In the low-pressure phase, the band gap slightly blue shift for the same reason than in YVO$_4$. At the phase transition the band gap decreases 1.1 eV. This is caused by an enhancement of the hybridization of O 2p orbitals with Cr 4s orbitals, which is the responsible of the pressure-driven transition and band-gap sharp decrease. The HP phase of CrVO$_4$ is also a direct gap material. Similar abrupt changes in the band-gap are observed at the pressure-induced transition of InVO$_4$ [106]. This is expected since both compounds have isomorphic low- and high-pressure phases.

In Fig. 13 we also show result of resistivity measurements in CrVO$_4$ [27]. This vanadate is a p-type semiconductor. Conductivity is extrinsic and it has been associated to the presence of impurities and defects in the crystal structure and therefore the conductivity (i.e. the resistivity) is little affected by pressure in the low-pressure phase. At the phase transition there is a drastic drop of the resistivity. The decrease of the resistivity has been related to an increase of the free-carrier concentration connected to the creation of oxygen vacancies at the phase transition [27]. No evidence of metallization has been observed in InVO$_4$ and CrVO$_4$. Indeed, results on vanadium dioxide indicate that pressure-induced metallization should occur in vanadates beyond 40 GPa [127]. The best candidate for it is FeVO$_4$, which has a band gap near 2 eV at ambient conditions. In this material electron–electron correlations could trigger metallization at much lower pressure than 40 GPa.

## VI. FUTURE PERSPECTIVES

In this section, we will present the different avenues that high pressure applied on vanadates can offer. We will first analyze magnetic properties. Several rare-earth vanadates are known to undergo magnetic transitions at low-temperature [67, 68]. Such transitions correlate with structural transitions which are related to Jahn-Teller distortions. Under low-temperature FeVO$_4$ also undergoes structural and magnetic changes, which evidence strong magneto-structural correlations [128]. The opposing effects of high pressure and the Jahn-Teller distortion led to many intriguing phenomena which are still not well understood [73, 129]. Pressure induces changes in magnetic behavior of compounds and the relation between magnetic and structural properties is also not fully known [130]. The study of $MVO_4$ vanadates at low-temperature could lead to important contributions to these two interesting subjects. Now-a-days both neutron diffraction [131] and single-crystal x-ray diffraction [132] can be performed under combined low-temperature and high pressure. Such studies would be crucial to deepen the knowledge on the behavior of $MVO_4$ vanadates under low-temperature and high-pressure. Another interesting subject to be studied under HP and low-temperature is the behavior of phonons. In particular, an anomalous behavior of the phonons of $TbVO_4$ has been reported to occur at low-temperature [133]. This phenomenon has been associated to an anharmonic phonon–phonon coupling, which could probably lead to a negative thermal expansion for $TbVO_4$ in the low-temperature limit. The combination of Raman and x-ray diffraction experiments at low-temperature and high-pressure could bring light to this interesting subject.

In addition to high-pressure low-temperature experiments, there is also a need of further high-pressure high-temperature experiments. These experiments are needed to determine accurate phase diagrams. Simultaneous HP-HT conditions have been used for the synthesis of metastable phases of vanadates [103, 134]. However, a systematic HP-HT study has been only carried out in $ErVO_4$ [135]. It has been reported that high-temperature lead to an anharmonic behavior of Raman modes [136]; which should have consequences in the specific heat and thermal expansion coefficients. To explore phonons and structural properties under HP-HT could lead therefore to important information for the better knowledge of $MVO_4$ vanadates.

Elastic constants of vanadates have provided information on the mechanism driving phase transitions [90, 137]. Studying them under compression, in particular in combination with magnetic fields, could be quite important for understanding magnetoelastic effects. In special, such studies would be helpful for the comprehension of the reported anomalies of Young modulus [138]. Unfortunately, experimental studies of elastic constant have been only carried out only at ambient conditions [57, 139, 140, 141, 142] and all the high-pressure information has been obtained by means of density-functional theory calculations [3]. The study of elastic properties under compression should be therefore the focus of future studies. High-quality single crystals of vanadates can be polished on different orientations and can be used to study elastic properties under high-pressure by ultrasonic measurements in a large-volume press [143] and by combined Brillouin spectroscopy and x-ray diffraction [144]. Such studies will bring light on transition mechanisms and mechanical properties of vanadates at high-pressure conditions.

Nanoparticles of $MVO_4$ materials have been mainly studied putting the focus on their size-dependent properties [145, 146]. It is known that particle size can significantly influence the high-pressure behavior of materials [147]. In particular, structural sequences and bulk modulus can be affected [148, 149], and new metastable phases can be recovered after pressure cycling [147, 148]. High-pressure studies on nanocrystal vanadates have been carried out in $HoVO_4$, $LaVO_4$, and $YVO_4$ among others [150 - 152]. In $LaVO_4$ and $HoVO_4$ nanocrystals behave similarly than their bulk counterparts. However, the situation of $YVO_4$ is different. In particular a structural transformation from the zircon-type structure to an amorphous phase has been found. This indicates that it is timely to carry our more studies on nanocrystal vanadates. Such studies could offer novel approaches for the engineering of original materials with tailored properties [152]. One of the issues that could be interesting to explore is the possibility of preparing phase of $InVO_4$ with hexa-coordinated vanadium. This could be probably obtained from nanoparticles prepared by calcination techniques an amorphous precursor [153]. Such nanoparticles doped with lanthanides could be very promising for biomedical applications [154, 155].

Another area that needs to be explored in the future is the behavior of all the vanadates that remain unstudied. One group of them are $SbVO_4$ [156], $RhVO_4$ [157], and $NbVO_4$ [158]. These vanadates crystallize in a disorder rutile-type structure with trivalent atoms and V occupying each of them randomly half of the cationic sites. The structure is shown in Fig. 14. Both cations are in octahedral coordination, being the structure described by space group $P4_2/mnm$ [158]. Long ago, Seifert has proposed that related rutile-type $InSbO_4$ undergoes a transition to the $\alpha$-$PbO_2$ structure at 11 GPa and 800ºC [159]. However, the existence of such a structure has never been studied in vanadates. It is well-known that rutile-type oxides undergo under compression subsequent transition to the $CaCl_2$-type and $\alpha$-$PbO_2$-type structures [160, 161]. Therefore, it would not be surprising that in $SbVO_4$, $RhVO_4$, and $NbVO_4$ similar transition would take place. Another possibility for these vanadates is the transformation at relative low pressure into zircon- or scheelite-type structure. Notice that the space group of rutile in $I4_1/amd$, which is a subgroup of the space group of rutile, $P4_2/mnm$. Indeed the zircons structure can be obtained by doubling the unit-cell of rutile along c-axis, and moving half of the cations (for instance Sb) in such a wave that they become dodecahedrically coordinated and the other half (V atoms) will become tetrahedrically coordinated. Thus both zircon and scheelite sound as good candidates for high-pressure structures of $SbVO_4$, $RhVO_4$, and $NbVO_4$ according to crystal-chemistry arguments.

In the case of $NbVO_4$, the situation is even more interesting, since according to the materials project there are many possible polymorphs for these compounds, which have similar formation energy [162]. The one with the lowest energy (i.e. the most stable) is a tetragonal structure described by space group $I-4m2$, in which both Nb and V atoms are in octahedral coordination. Interestingly this structure is predicted to be ferromagnetic with a magnetic moment of 1.987 μB and metallic. In addition, it is expected to be very uncompressible with a bulk modulus larger than 220 GPa [108]. A final interesting fact of $NbVO_4$ is that it is a compound where both octahedral are "suffering" a Jahn-Teller distortion [163, 164]. All the facts, described in this and the previous paragraph, indicate that $NbVO_4$ is a very interesting candidate for future high-pressure studies.

In addition, to the three compounds discussed above there other orthovanadates of interest to study under high pressure. They include $TaVO_4$, $TiVO_4$, $TlVO_4$ ($CrVO4$), $AlVO_4$ ($FeVO4$), $VPO_4$, and $TeVO_4$ [165 – 168]. $TlVO_4$ is expected to behave under compression in a similar way that $InVO_4$ and $CrVO_4$, $AlVO_4$ is isostructural to $FeVO_4$ and therefore should also have a very rich high-pressure polymorphism. $TeVO_4$ is of interest because of its magnetic properties [167]. In analogy with $GaPO_4$ [169] which has very interesting piezoelectric properties, it could be fascinating to study the high-pressure behavior of $GaVO_4$. We would also like to draw the attention to $BVO_4$ which could be an ultra-hard material. However, both vanadates have not been accessible up to know. The structure of them can be predicted evolutionary crystal prediction methods [170]. On top of that, machine learning techniques can be used to design synthesis methods [171]. The combination of these novel techniques combined with high-pressure studies could open very promising opportunities for the study of MVO$_4$ vanadates under pressure.

To conclude we would like to add that the study of orthovanadates under compression has been contributing to the understanding of the high-pressure behavior of other vanadates, like zinc pyrovanadate and related compounds [172]. In particular, the systematics established for structural sequences on MVO$_4$ compounds has been useful for understanding the structural sequence, mechanical properties, and electronic properties of other vanadates. They can be also used as a guide for the studied of many other vanadates not studied in deep under high-pressure; including perovskite-type vanadates (e.g. LaVO$_3$) [173], spinel-type vanadates (e.g. LiV$_2$O$_4$) [174, 175], delafossite-type vanadates (e.g. LiVO$_2$) [176], and metavanadates (e. g. Ca$_3$(VO$_4$)$_2$ [177]). All these compounds have practical applications and his study under compression could lead to important discoveries. The study of spinel-type vanadates could be of particular interest given that it has been reported anomalous phenomena in them [178]. For instance, a sign change has been found in the pressure dependence of the Néel temperature, and a bulk modulus much smaller than in other spinel-type compounds have been measured.

## VII. CONCLUDING REMARKS

The application of high-pressure could lead to many novel and interesting phenomena. In the case of MVO$_4$ orthovanadates, within a limited pressure range, structural phase transitions have been induced by pressure in many different vanadates. The most remarkable case is FeVO$_4$, which below 6 GPa undergoes three phase transitions. These phase transitions bring associated to them drastic changes in the vibrational, mechanical, and electronic properties of the vanadates. One of the most noticeably phenomena is the collapse of the electronic band-gap found in many vanadates in parallel with the occurrence of a structural phase transition. This abrupt change of the band-gap can be larger than 1 eV in compounds like InVO$_4$ and CrVO$_4$. In this perspective article, we report the most recent findings in rare-earth vanadates, and other MVO$_4$ vanadates. Phase transitions are described and the high-pressure behavior of physical properties described. A systematic understanding of all the reported results is presented. In particular, we have explained why pressure allows modifications of these materials putting emphasis in changes that could help for the developing of their many technological applications. Finally, we dedicated a complete section to present interesting unstudied problems, suggesting different avenues that would be interesting to explore in the following years. We hope this article will trigger future studies leading to interesting discoveries in fundamental physics and applications.

**ACKNOWLEDGMENTS**

The author thanks the financial support given by the Spanish Ministry of Science, Innovation and Universities under grants MAT2016-75586-C4-1-P, PID2019-106383GB-C41, and RED2018-102612-T (MALTA Consolider-Team network) and by Generalitat Valenciana under grant Prometeo/2018/123 (EFIMAT). The author is grateful to all the collaborators with whom they have the pleasure to work together on the study of vanadates under extreme conditions.

**DATA AVAILABILITY STATEMENT**

The data that support the findings of this study are available from the corresponding author upon reasonable request.

**Figure 1:** Different perspective of the crystal structure of zircon-type (a), scheelite-type (b), monazite-type (c), and BaWO$_4$-II-type (d) structures. Vanadium (rare-earth) coordination polyhedra are shown in red (green-blue). The small red circles are the oxygen atoms.

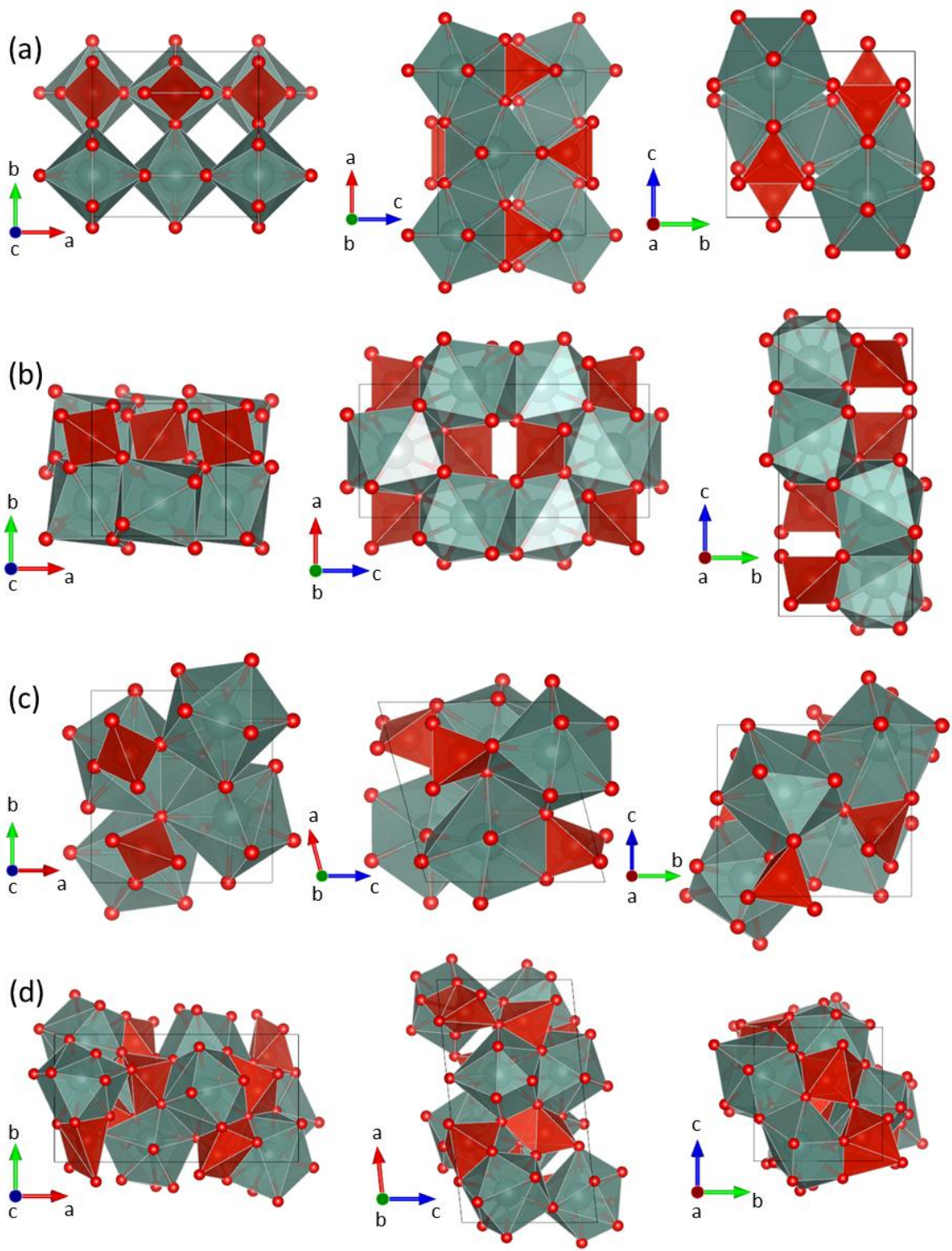

**Figure 2:** Schematic representation of the high-pressure structural sequence of alkaline-earth vanadate.

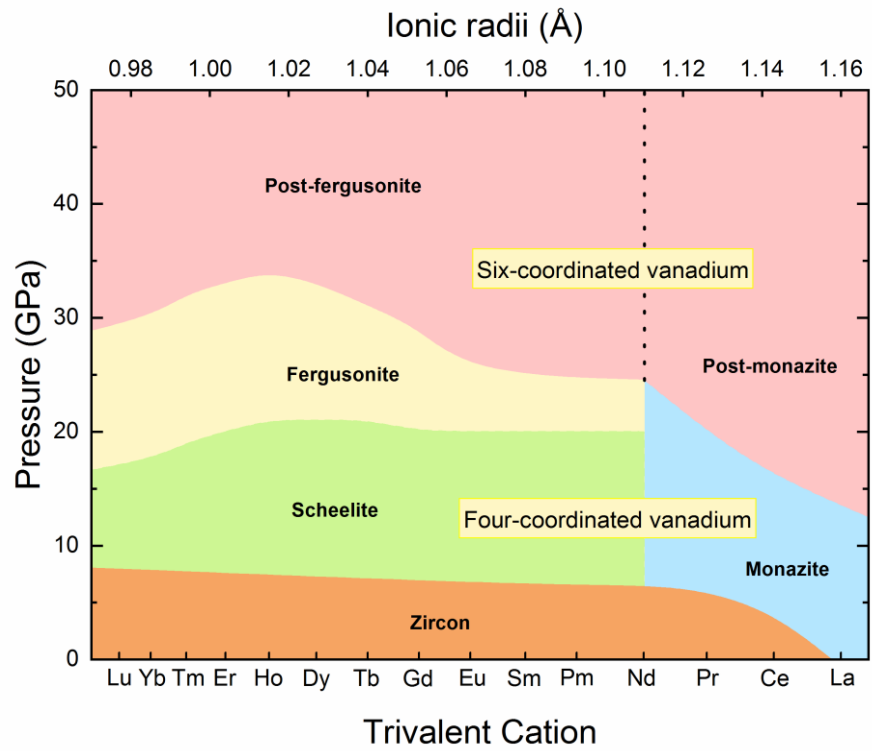

**Figure 3:** Powder XRD patterns measured in TbVO$_4$ at ambient pressure (low-pressure phase, space group I4$_1$/amd)) and 8.8 GPa (high-pressure phase, space group I4$_1$/a) illustrating the zircon-scheelite transition [41]. In addition to the experimental results, we show the residuals and ticks indicating positions of diffraction peaks. Structures are identified by its space group.

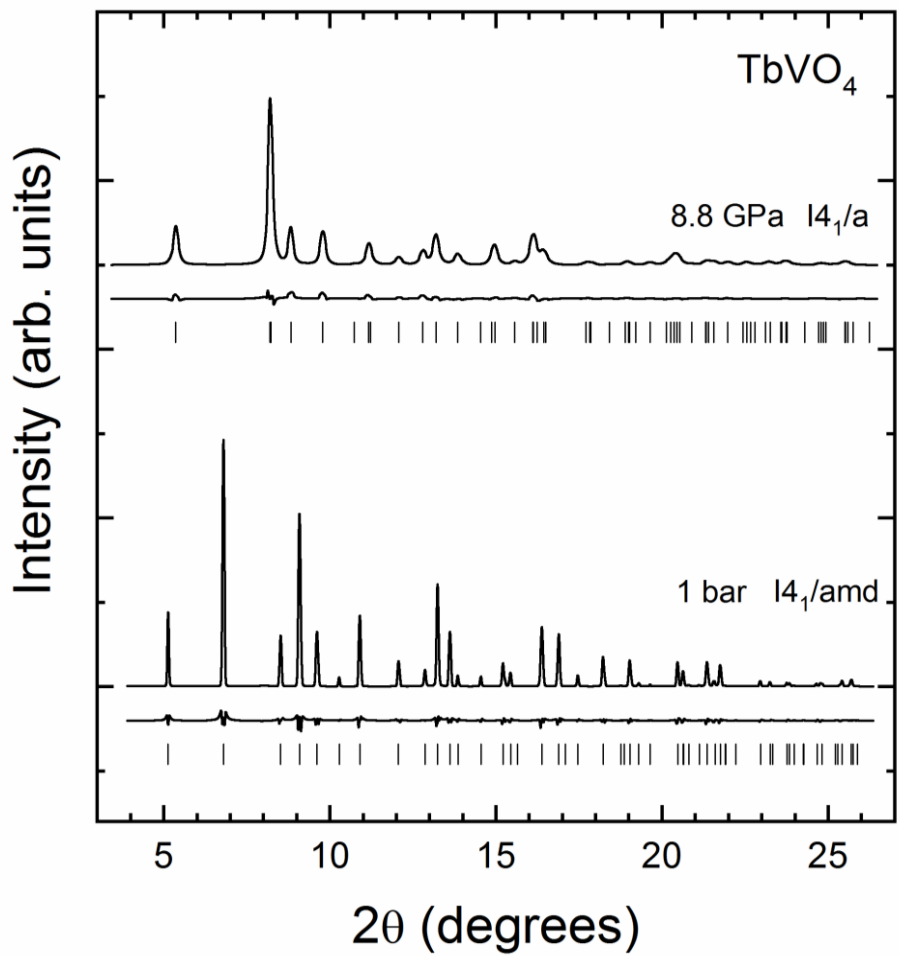

**Figure 4**: Powder XRD patterns measured in CeVO$_4$ at 1.7 GPa (low-pressure phase, space group I4$_1$/amd) and 5.6 GPa (high-pressure phase, space group P2$_1$/n) illustrating the zircon-monazite transition [41]. In addition to the experimental results, we show the residuals and ticks indicating positions of diffraction peaks. Structures are identified by its space group.

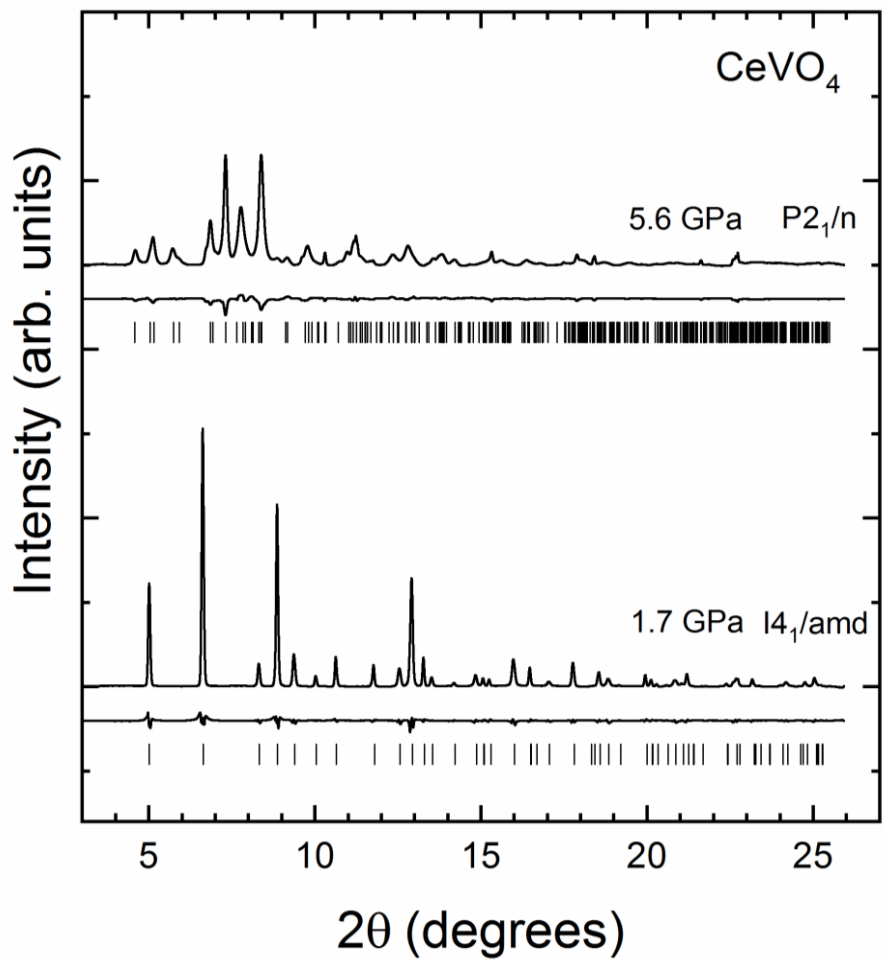

**Figure 5:** Raman spectra of TbVO$_4$ illustrating the changes induced by phase transitions [59]. We show the spectrum of different phases using different colors. The phase and pressure are indicated.

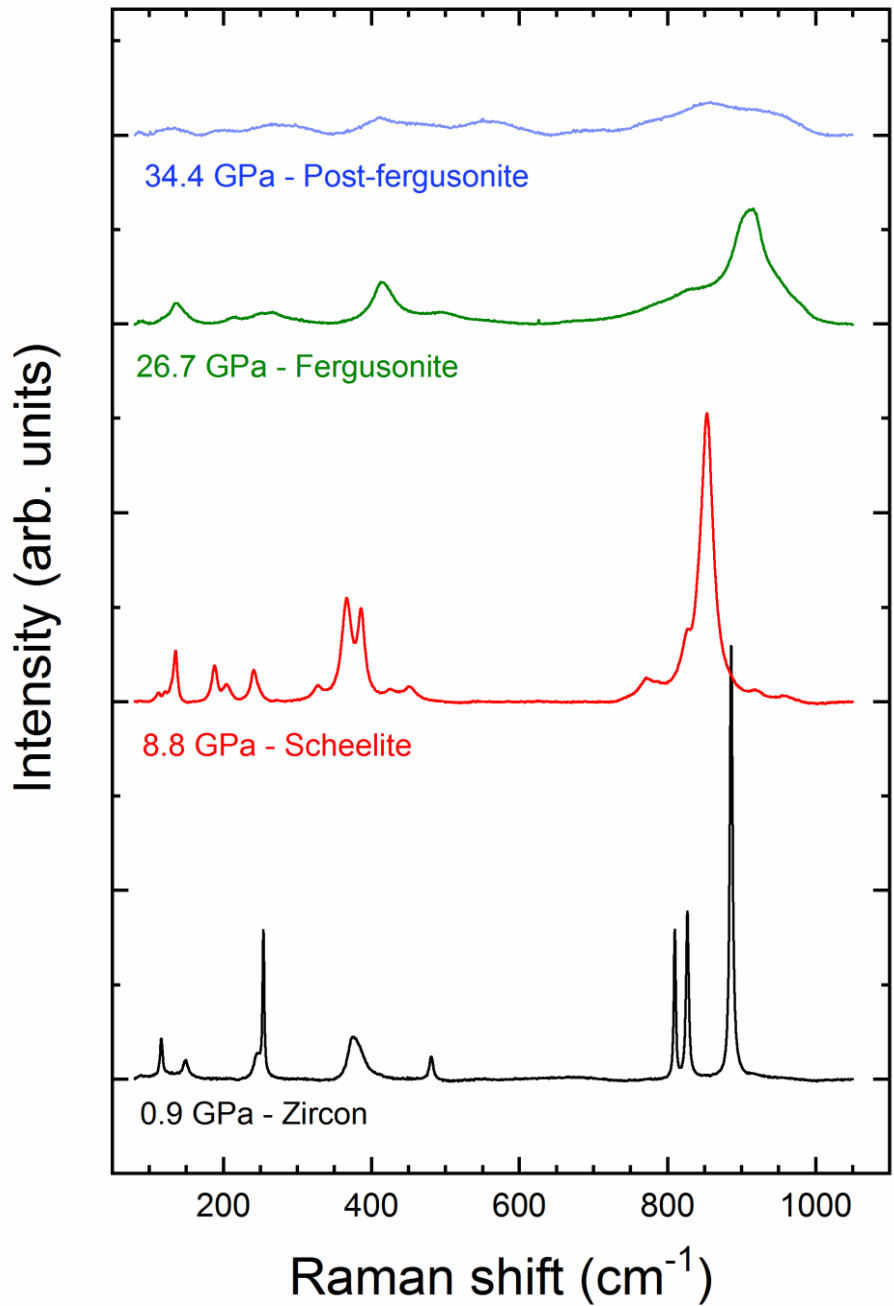

**Figure 6:** Pressure dependence of the Raman modes measured in zircon-type and scheelite-type $TbVO_4$ [59]. Circles represent the experimental results and lines the results of calculations. We only show the modes observed in the experiments. The modes softening under compression are shown in red. Modes are labeled with their symmetry assignment.

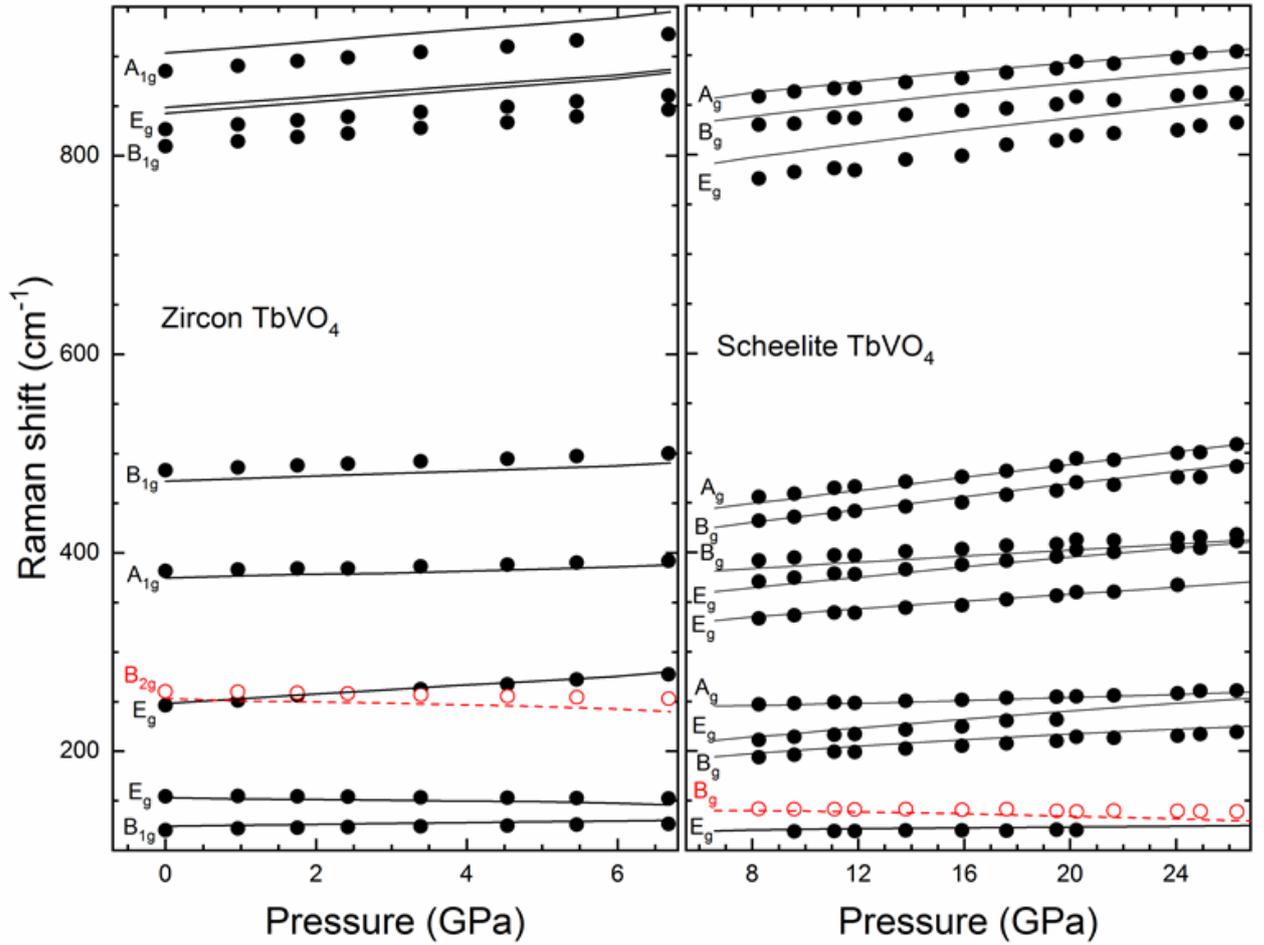

**Figure 7:** Photoluminescence spectra of zircon $Sm_xNd_{1-x}VO_4$ ($\lambda_{exc}$ = 470 nm) [23]. Concentrations of Sm (x) are indicated.

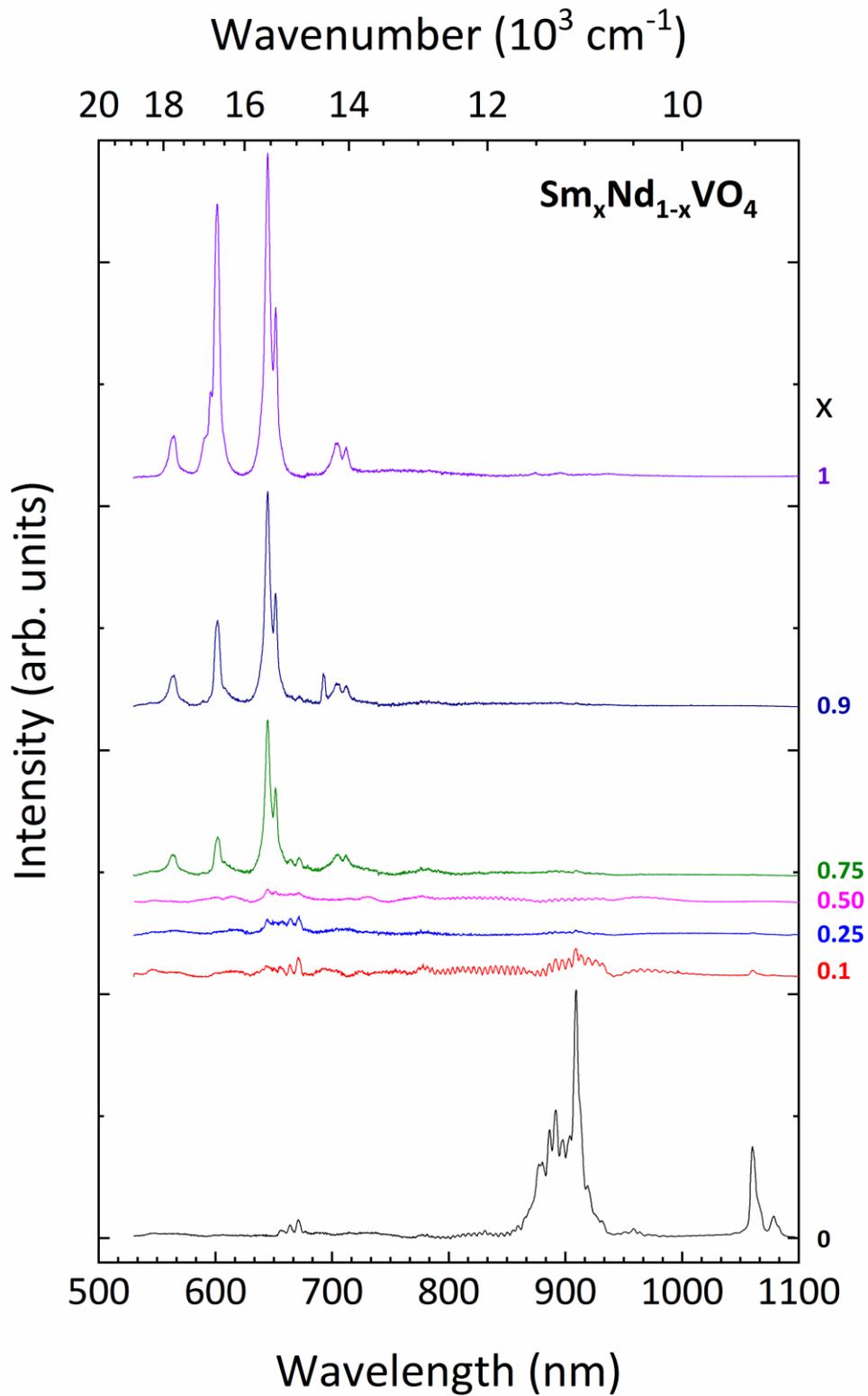

**Figure 8:** Refinements for XRD patterns measured in Sm$_{0.5}$Nd$_{1-x}$VO$_4$ at different pressures [21]. (a) zircon structure at 2.9 GPa; (b) mixture of scheelite and monazite at 12.0 GPa; (c) fergusonite at 15.5 GPa. In all plots, black and red ticks indicated the peaks of the different phases of Sm$_{0.5}$Nd$_{1-x}$VO$_4$ and green ticks indicate the peaks due to copper. The circles are the experiments, the red lines the fits, and the blue lines the residuals.

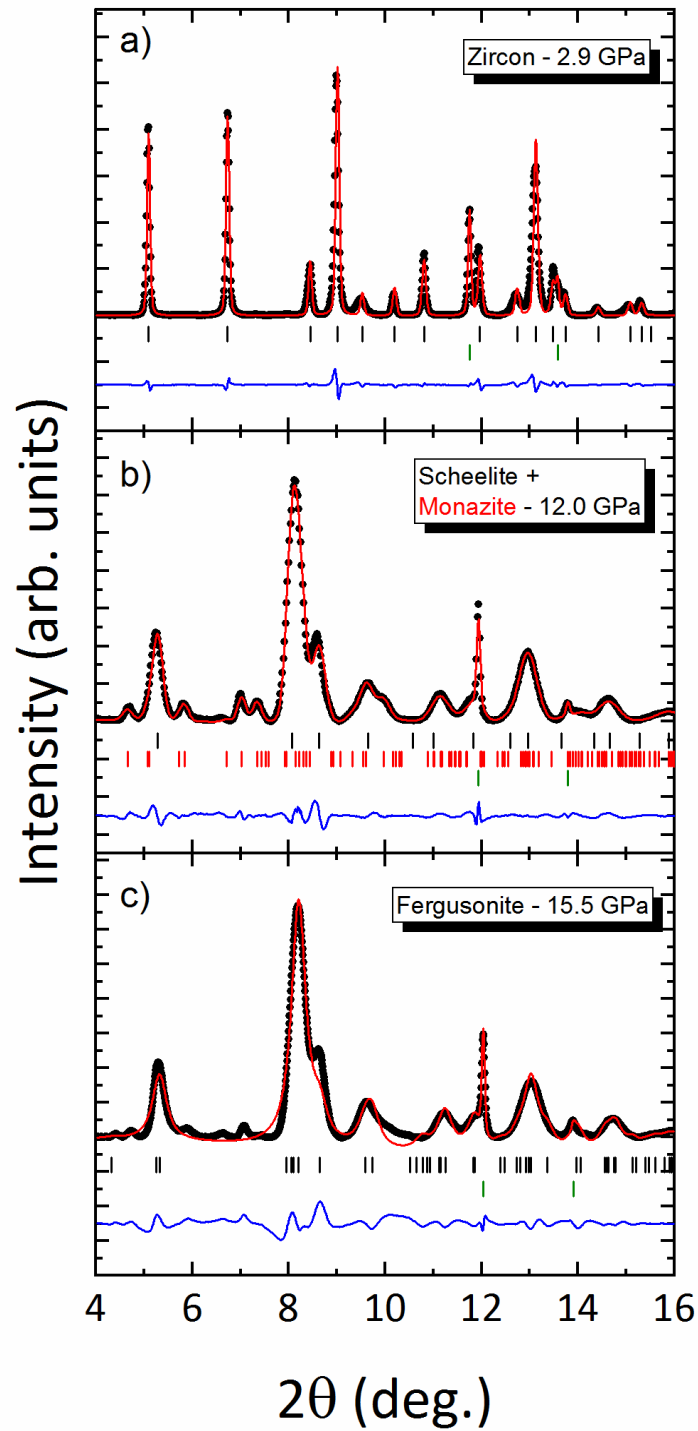

**Figure 9:** Different perspective of the crystal structure of CrVO$_4$ (top) and FeVO$_4$ (bottom) structures. Vanadium (chromium/iron) coordination polyhedra are shown in red (blue/brown). The small red circles are the oxygen atoms.

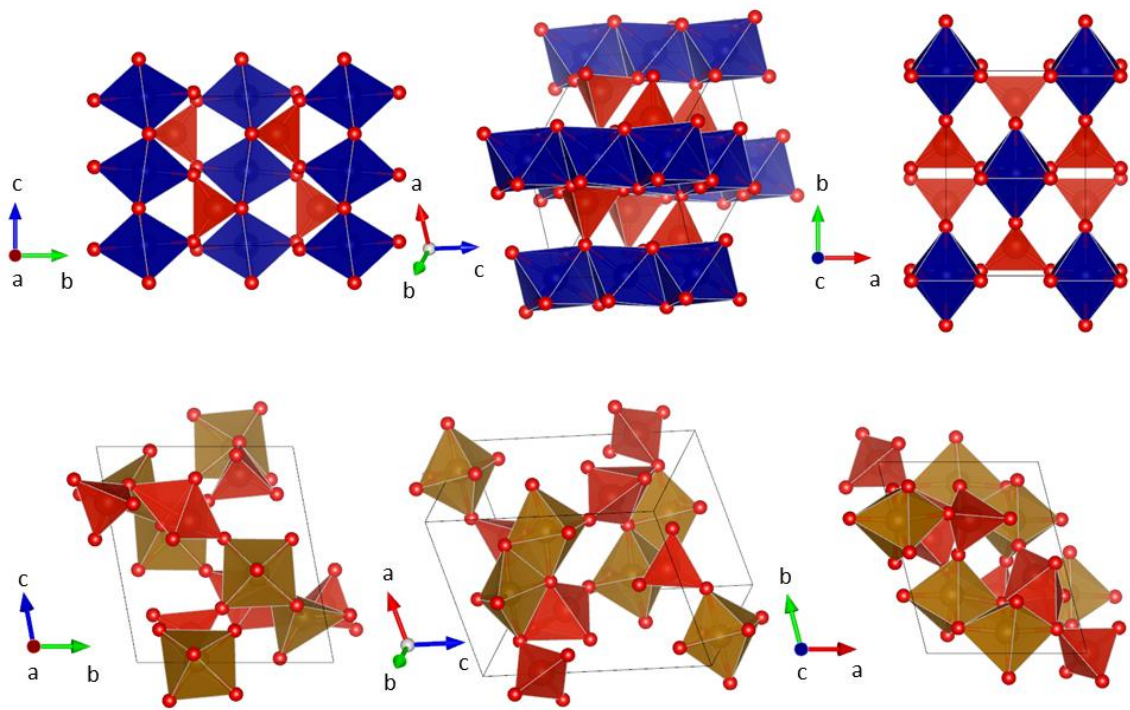

**Figure 10:** (Left) XRD patterns of the low-pressure and high-pressure phases of InVO$_4$ [26]. Symbols are experiments. Black solid lines are Rietveld refinements. Red solid lines are residuals. Ticks indicate the positions of Bragg peaks (Right) Perspective of the crystal structure of both phases.

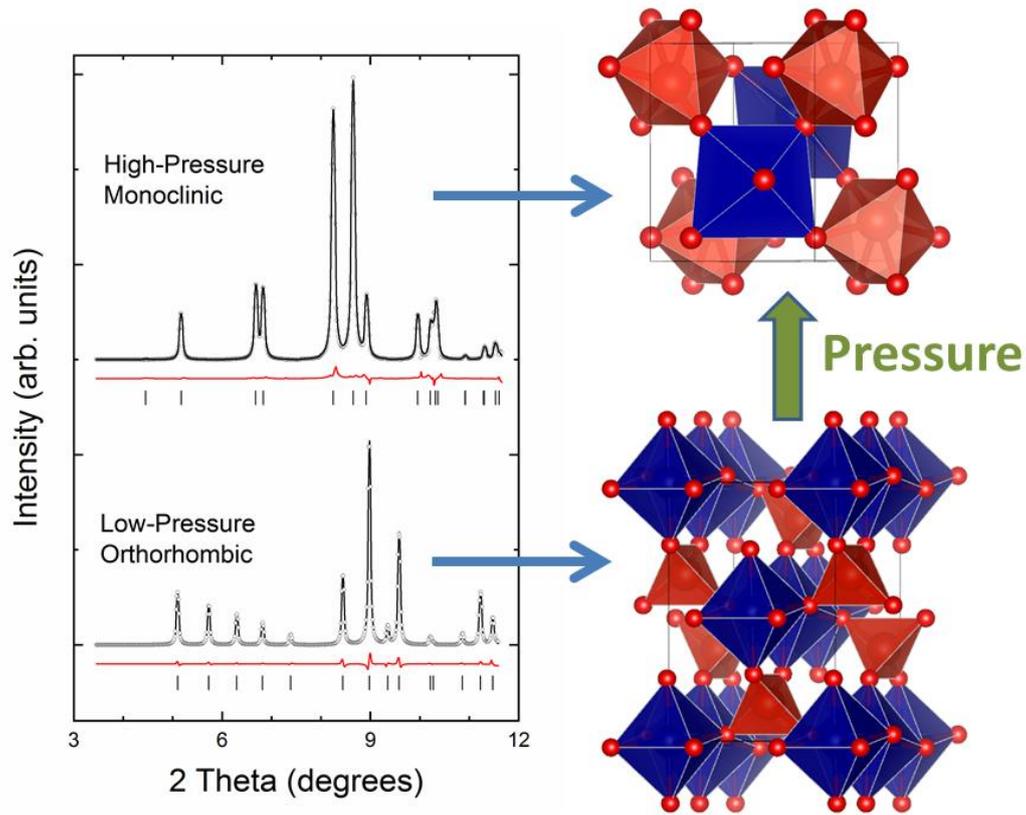

**Figure 11:** Schematic representation of the band structure of rare-earth vanadates.

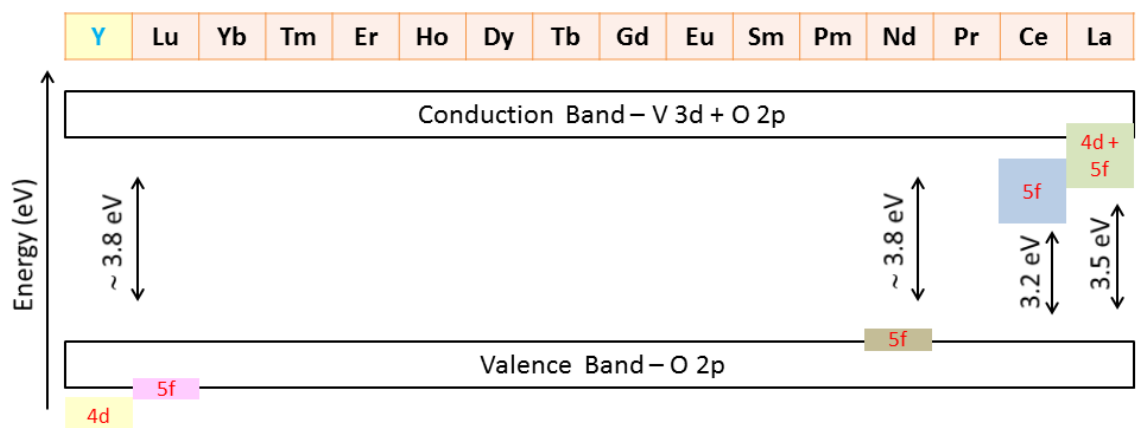

**Figure 12:** (Top left) Optical absorption spectra of LaVO$_4$ at different pressures [15, 116]. (Top right) Optical absorption spectra of YVO$_4$ at different pressures. (Bottom left) Pressure dependence of the band gap energy (Eg) in LaVO$_4$. Symbols are experiments and lines linear fits. (Bottom right) Pressure dependence of the band gap energy (Eg) in YVO$_4$. Symbols are experiments and lines linear fits.

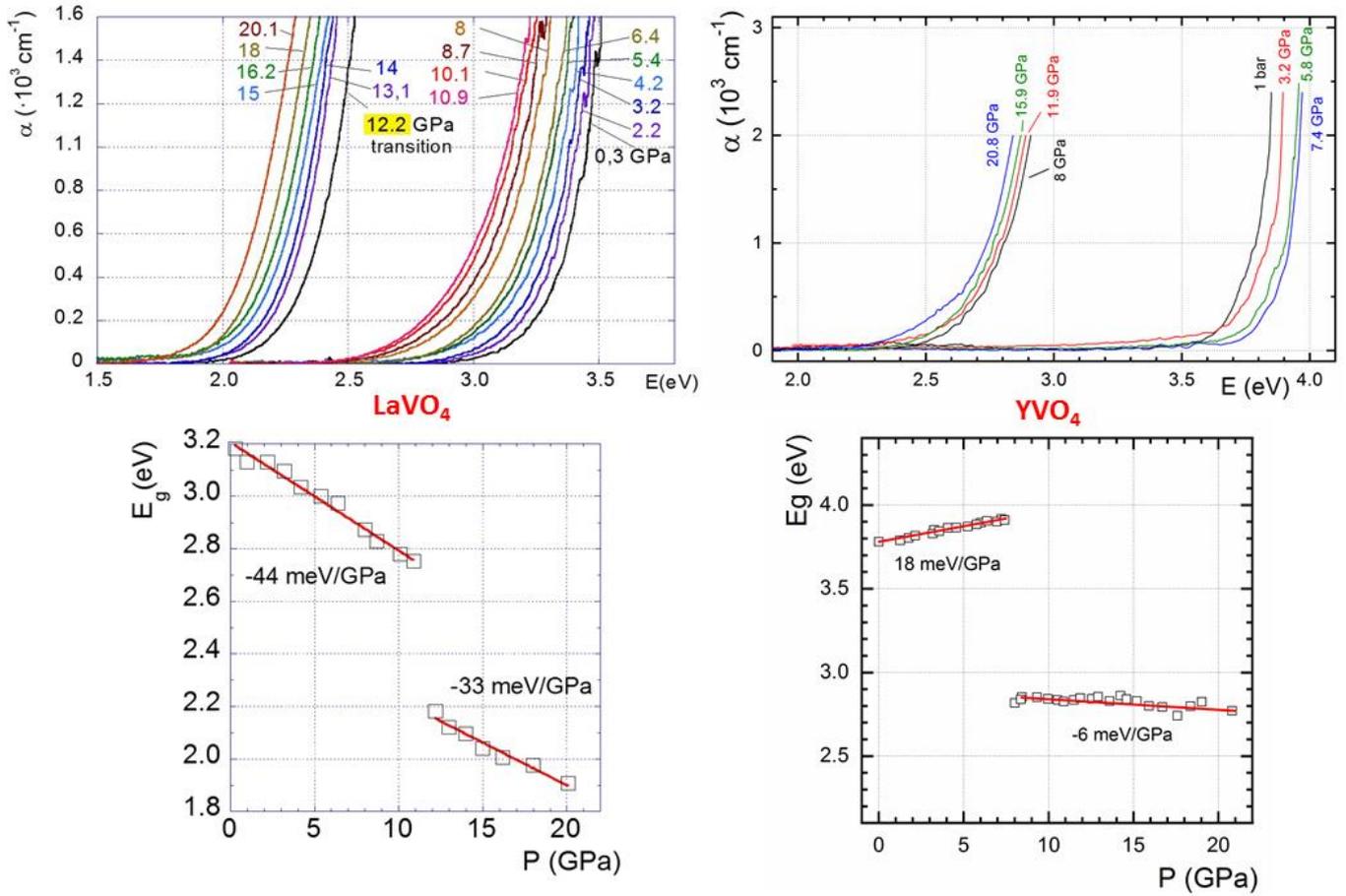

**Figure 13:** (Left) Pressure dependence of the band gap energy (Eg) in CrVO$_4$ [26[. Symbols are experiments and lines results from DFT calculations. For the low-pressure phase a linear fit to experiments is also included (Right) Pressure dependence of resistivity in CrVO$_4$ [26].

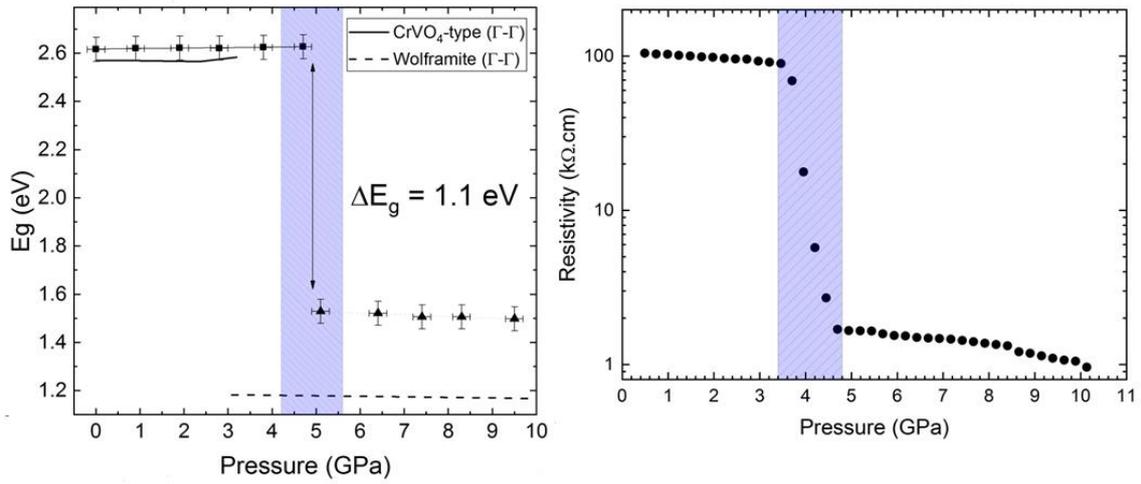

**Figure 14:** Rutile-type structure of SbVO$_4$, with Sb and V randomly occupying the octahedral positions (with a partial occupation 0.5 for each atom).

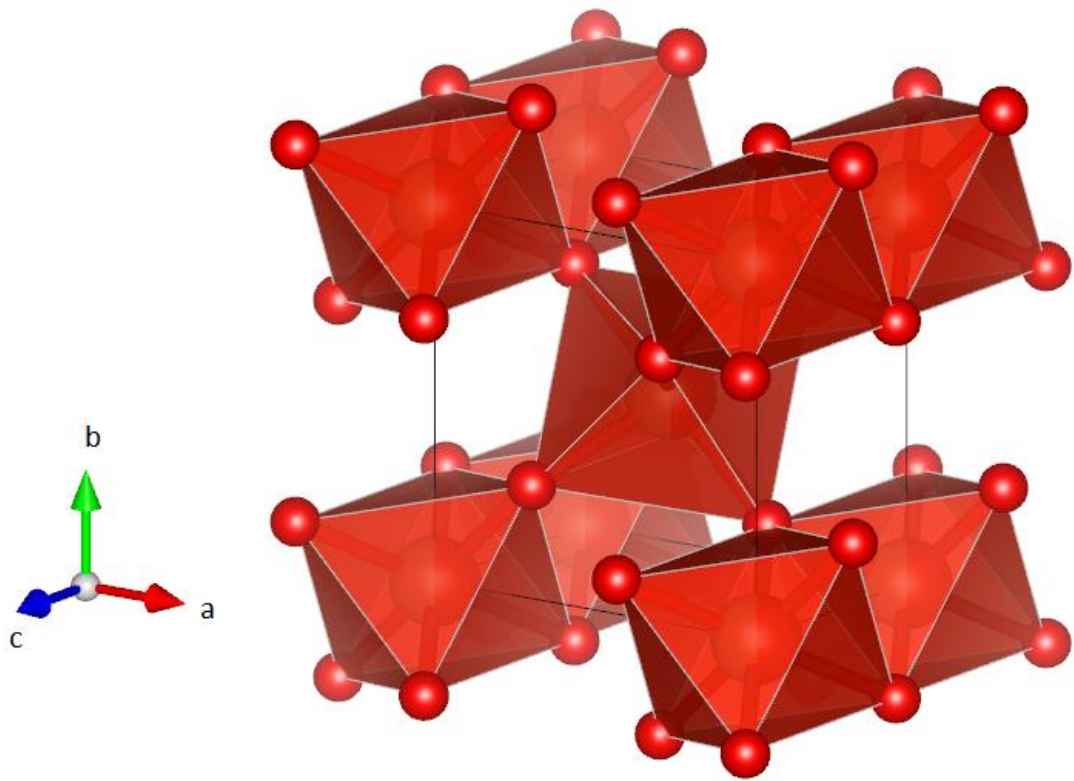

**Figure 15:** tetragonal structure of NbVO4 propsed by density-functional theory calculations. SbO$_6$ and VO$_6$ octahedra are shown in green and red respectively.

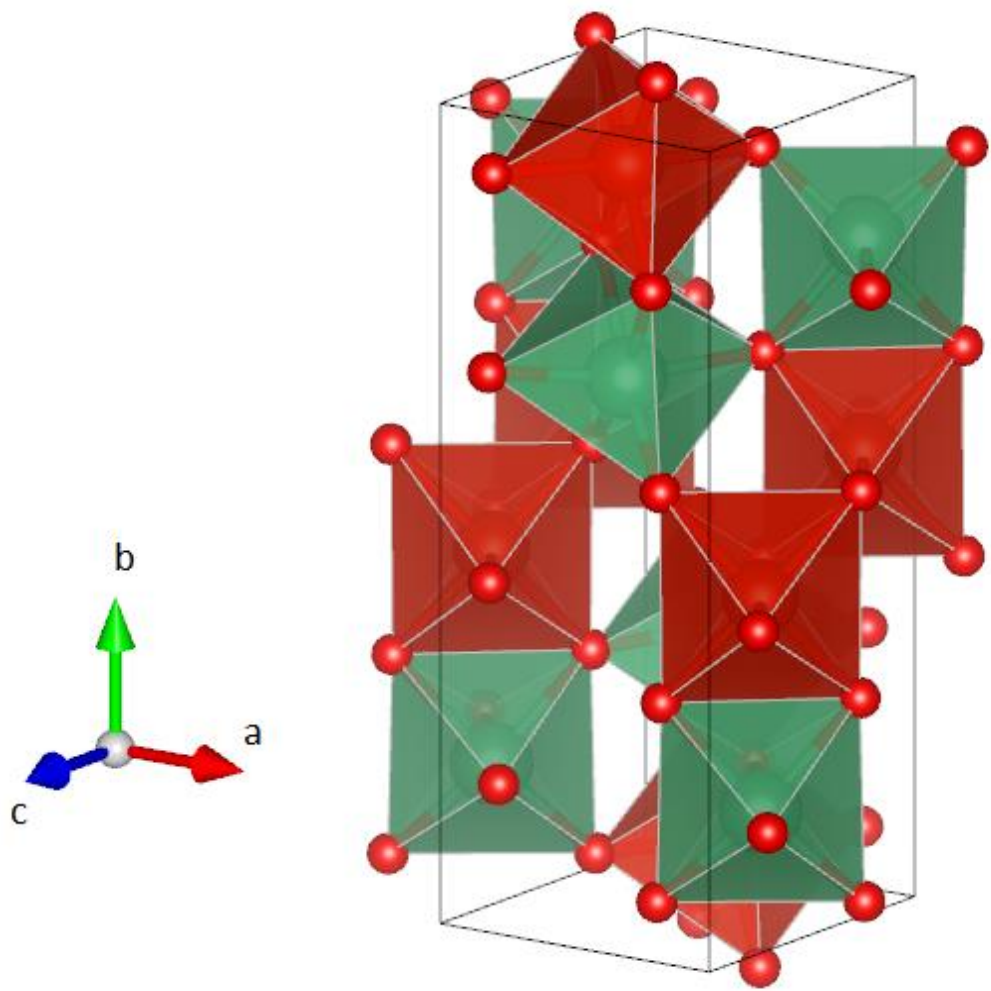